\documentstyle[12pt]{article}

\def\hybrid{\topmargin -20pt	\oddsidemargin 0pt
	\headheight 0pt	\headsep 0pt
	\textwidth 6.25in	
	\textheight 9.5in	
	\marginparwidth .875in
	\parskip 5pt plus 1pt	\jot = 1.5ex}

\hybrid

\def\baselinestretch{1.2}

\catcode`\@=11

\def\marginnote#1{}
\newcount\hour
\newcount\minute
\newtoks\amorpm
\hour=\time\divide\hour by60
\minute=\time{\multiply\hour by60 \global\advance\minute by-\hour}
\edef\standardtime{{\ifnum\hour<12 \global\amorpm={am}%
	\else\global\amorpm={pm}\advance\hour by-12 \fi
	\ifnum\hour=0 \hour=12 \fi
	\number\hour:\ifnum\minute<10 0\fi\number\minute\the\amorpm}}
\edef\militarytime{\number\hour:\ifnum\minute<10 0\fi\number\minute}

\def\draftlabel#1{{\@bsphack\if@filesw {\let\thepage\relax
   \xdef\@gtempa{\write\@auxout{\string
      \newlabel{#1}{{\@currentlabel}{\thepage}}}}}\@gtempa
   \if@nobreak \ifvmode\nobreak\fi\fi\fi\@esphack}
	\gdef\@eqnlabel{#1}}
\def\@eqnlabel{}
\def\@vacuum{}
\def\draftmarginnote#1{\marginpar{\raggedright\scriptsize\tt#1}}

\def\draft{\oddsidemargin -.5truein
	\def\@oddfoot{\sl preliminary draft \hfil
	\rm\thepage\hfil\sl\today\quad\militarytime}
	\let\@evenfoot\@oddfoot	\overfullrule 3pt
	\let\label=\draftlabel
	\let\marginnote=\draftmarginnote
   \def\@eqnnum{(\theequation)\rlap{\kern\marginparsep\tt\@eqnlabel}%
\global\let\@eqnlabel\@vacuum}  }

\def\preprint{\twocolumn\sloppy\flushbottom\parindent 2em
	\leftmargini 2em\leftmarginv .5em\leftmarginvi .5em
	\oddsidemargin -.5in	\evensidemargin -.5in
	\columnsep .4in	\footheight 0pt
	\textwidth 10.in	\topmargin  -.4in
	\headheight 12pt \topskip .4in
	\textheight 6.9in \footskip 0pt
	\def\@oddhead{\thepage\hfil\addtocounter{page}{1}\thepage}
	\let\@evenhead\@oddhead	\def\@oddfoot{}	\def\@evenfoot{} }

\def\numberbysection{\@addtoreset{equation}{section}
	\def\theequation{\thesection.\arabic{equation}}}

\def\underline#1{\relax\ifmmode\@@underline#1\else
	$\@@underline{\hbox{#1}}$\relax\fi}

\def\titlepage{\@restonecolfalse\if@twocolumn\@restonecoltrue\onecolumn
     \else \newpage \fi \thispagestyle{empty}\c@page\z@
	\def\thefootnote{\fnsymbol{footnote}} }

\def\endtitlepage{\if@restonecol\twocolumn \else \newpage \fi
	\def\thefootnote{\arabic{footnote}}
	\setcounter{footnote}{0}}  

\catcode`@=12
\relax

\def\figcap{\section*{Figure Captions\markboth
	{FIGURECAPTIONS}{FIGURECAPTIONS}}\list
	{Figure \arabic{enumi}:\hfill}{\settowidth\labelwidth{Figure
999:}
	\leftmargin\labelwidth
	\advance\leftmargin\labelsep\usecounter{enumi}}}
 \relax
\def\tablecap{\section*{Table Captions\markboth
	{TABLECAPTIONS}{TABLECAPTIONS}}\list
	{Table \arabic{enumi}:\hfill}{\settowidth\labelwidth{Table
999:}
	\leftmargin\labelwidth
	\advance\leftmargin\labelsep\usecounter{enumi}}}
 \relax
\def\reflist{\section*{References\markboth
	{REFLIST}{REFLIST}}\list
	{[\arabic{enumi}]\hfill}{\settowidth\labelwidth{[999]}
	\leftmargin\labelwidth
	\advance\leftmargin\labelsep\usecounter{enumi}}}
 \relax
\makeatletter
\newcounter{pubctr}
\def\publist{\@ifnextchar[{\@publist}{\@@publist}}
\def\@publist[#1]{\list
	{[\arabic{pubctr}]\hfill}{\settowidth\labelwidth{[999]}
	\leftmargin\labelwidth
	\advance\leftmargin\labelsep
	\@nmbrlisttrue\def\@listctr{pubctr}
	\setcounter{pubctr}{#1}\addtocounter{pubctr}{-1}}}
\def\@@publist{\list
	{[\arabic{pubctr}]\hfill}{\settowidth\labelwidth{[999]}
	\leftmargin\labelwidth
	\advance\leftmargin\labelsep
	\@nmbrlisttrue\def\@listctr{pubctr}}}
 \relax
\makeatother
\newskip\humongous \humongous=0pt plus 1000pt minus 1000pt

\newif\ifdtup

\relax

\def\s{\sigma}
\def\thefootnote{\fnsymbol{footnote}}
\def\be{\begin{equation}}
\def\ee{\end{equation}}
\def\ba{\begin{eqnarray}}
\def\ea{\end{eqnarray}}
\def\d{\partial}

\def\G{\Gamma}
\def\S{\Sigma}
\def\s{\sigma}

\def\a{\alpha}
\def\th{\theta}

\def\r{\rho}
\def\l{\lambda}

\def\e{\epsilon}

\def\a{\alpha}
\def\b{\beta}
\def\vphi{\varphi}
\def\ha{{1\over 2}}

\begin{document}

\renewcommand{\theequation}{\arabic{equation}}
\newcommand{\beq}{\begin{equation}}
\newcommand{\eeq}[1]{\label{#1}\end{equation}}
\newcommand{\ber}{\begin{eqnarray}}
\newcommand{\eer}[1]{\label{#1}\end{eqnarray}}
\begin{titlepage}
\begin{center}

\hfill CERN--TH/96--15\\
\hfill ENSLAPP--A--578/96\\
\hfill THU--96/04\\
\hfill hep--th/9601158\\

\vskip .3in

{\large \bf STRING EFFECTS AND FIELD THEORY PUZZLES \\
WITH SUPERSYMMETRY}
\footnote{Contribution to the proceedings of the 5th Hellenic School
and Workshops on Elementary Particle Physics, Corfu, 3-24 September
1995}
\vskip 0.3in

{\bf Ioannis Bakas}
\footnote{Permanent address: Department of Physics, University of
Patras,
26110 Patras, Greece}
\footnote{e--mail address: BAKAS@SURYA11.CERN.CH,
BAKAS@LAPPHP8.IN2P3.FR}\\
\vskip .1in

{\em Theory Division, CERN, 1211 Geneva 23, Switzerland, and\\
Laboratoire de Physique Theorique ENSLAPP, 74941 Annecy-le-Vieux,
France}
\footnote{Present address}\\

\vskip .3in

{\bf Konstadinos Sfetsos}
\footnote{e--mail address: SFETSOS@FYS.RUU.NL}\\
\vskip .1in

{\em Institute for Theoretical Physics, Utrecht University\\
     Princetonplein 5, TA 3508, The Netherlands}\\

\vskip .1in

\end{center}

\vskip .3in

\begin{center} {\bf ABSTRACT } \end{center}
\begin{quotation}\noindent
We investigate field theory puzzles occuring in the interplay
between supersymmetry and duality in the presense of rotational
isometries (also known as non--triholomorphic in hyper--Kahler
geometry). We show that T--duality is always compatible with
supersymmetry, provided that non--local world--sheet effects are
properly taken into account. The underlying superconformal
algebra remains the same, and T--duality simply relates
local with non--local realizations of it. The non--local
realizations have a natural description using parafermion
variables of the corresponding conformal field theory. We also
comment on the relevance of these ideas to a possible
resolution of long standing problems in the quantum theory
of black holes.
\end{quotation}
\vskip .3cm
January 1996\\
\end{titlepage}
\vfill
\eject

\def\baselinestretch{1.2}
\baselineskip 16 pt

\subsection*{ Introduction }

Attempts to describe phenomena using inadequate theories usually lead
to
paradoxes. A prototype example in elementary particle physics is the
Klein
paradox, which ceases to exist once field theory replaces quantum
mechanics.
It is our purpose in this note (based on two lectures given by the
authors
in Corfu) to describe a paradox of the
effective field theory that occurs in the interplay between duality
and
supersymmetry, which on the other hand has a natural explanation
within string theory. We will see that certain aspects of our
investigation
could also be relevant to string phenomenology, in connection with
supersymmetry breaking mechanisms, which are usually considered at
the
level of the lowest order effective theory.
Moreover, we will suggest at the very end a string theoretical
framework for
providing a plausible resolution to long standing problems in the
quantum
theory of black holes, using non--locally realized superconformal
algebras.

The classical propagation of strings in a general target space with
metric $G_{\mu \nu}(X)$ and antisymmetric tensor field $B_{\mu
\nu}(X)$ is
described by the 2--dim $\sigma$--model Lagrangian density
${\cal L} = Q^{+}_{\mu \nu} \d_+ X^{\mu} \d_- X^{\nu}$, where
$Q^{+}_{\mu\nu}\equiv G_{\mu\nu} + B_{\mu\nu}$; we also introduce
$Q^{-}_{\mu\nu}\equiv G_{\mu\nu} - B_{\mu\nu}$ for later use.
The natural time coordinate on
the world--sheet is $\tau= \s^+ + \s^- $, while $\s= \s^+ - \s^-$
denotes the corresponding spatial variable.
We consider string backgrounds with a Killing symmetry associated
to a vector field $\d/{\d X^0}$, and denote the rest of the
target space coordinates by $X^i$, $i=1,\dots , d-1$.
We choose to work for convenience in an
adapted coordinate system where the background fields
are all independent of $X^0$.
It was shown by Buscher some time ago \cite{BUSCHER}
that under these circumstances a
dual $\s$--model can be found, as a solution of the $\beta$--function
equations, with background fields
\beq
\tilde G_{00} = {1\over G_{00}}~ ,~~~~ \tilde Q^\pm_{0i} =
\pm {Q^\pm_{0i}\over G_{00}}~ ,~~~~ \tilde Q^+_{ij}= Q^+_{ij} ~ - ~
{Q^+_{i0} Q^+_{0j} \over G_{00}}~,
\label{dualbou}
\end{equation}
while conformal invariance at 1--loop also requires
a shift in the dilaton field
$\Phi$ by $\ln (G_{00})$.

We wish to investigate how the supersymmetric
properties of $\s$-models behave
under T--duality. This will turn out to be an intriguing problem,
because
geometrical objects in the target space (like Kahler forms and
Killing spinors)
are not necessarily independent
of the Killing coordinate $X^0$, although the corresponding
$\s$--model
background fields always are. To achieve our goal we have to find a
direct
way to formulate the action of T--duality transformations on the
target
space coordinates themselves, from which the transformation
properties of all
other quantities in the target space can be easily deduced. We find
that the
relevant formulation here is the description of Abelian T--duality as
a canonical
transformation on the
world--sheet.
In this approach \cite{AALcan} one performs the transformation
$(X^0,P_0)\to
(\tilde X^0,\tilde P_0)$, defined by $\tilde P_0=\d_\s X^0$, $\d_\s
\tilde X^0 =
P_0$, where $P_0$ is the conjugate momentum to $X^0$ in the 2--dim
$\s$--model.
This amounts to a non--local redefinition of the
target space variable associated with the Killing symmetry,
\beq
X^0=\int \tilde Q^+_{\mu 0} \d_+ \tilde X^\mu d \s^+ ~
- ~ \tilde Q^-_{\mu 0} \d_- \tilde X^\mu d\s^- ~ ,
\label{rednonl}
\end{equation}
which in the Hamiltonian description of the 2--dim $\s$--model yields
(\ref{dualbou}); for earlier work on this subject see \cite{CuZa}.
Therefore, despite non--localities, the dual background fields are
locally related to the original ones by (\ref{dualbou}).
However, other geometrical
objects in target space that can generically depend on $X^0$ will
become
non--local in the dual face of the theory \cite{basfe}.
The canonical formalism offers the
advantage to explore explicitly the transformation properties of such
objects,
including Kahler forms in particular.

When are we faced with a paradox in the effective field theory
approach to
quantum string dynamics? Only when we insist on having a consistent
description
in terms of local effective field theories and forget various
non--local
world--sheet effects generated by (\ref{rednonl}),
when it is appropriate. Such non--local effects constitute
the main theme of our work, and as we will see next by studying the
interplay
between T--duality and supersymmetry, they have a crucial role in
understanding the string resolution to paradoxes of the effective
theory.

\subsection*{ Duality and world--sheet supersymmetry: generalities }

It is well known that any background can be made $N=1$ supersymmetric
on the world--sheet \cite{FRTO}, and
there cannot possibly be a clash with duality (Abelian and
non--Abelian as
well as $S$--duality) in this case.
In contrast, extended $N=2$ supersymmetry \cite{zumino,ALFR}
requires that the background is such that an (almost) complex
(hermitian)
structure $F^\pm_{\mu\nu}$ exists in each sector associated to the
right and
left-handed fermions.
Similarly, $N=4$ extended supersymmetry \cite{ALFR,HSPB}
requires that there exist three independent complex structures in
each sector,
$(F_I^\pm)_{\mu\nu}$, $I=1,2,3$.
The complex structures are covariantly constant with respect to
generalized
connections that include the torsion,
they are represented by antisymmetric matrices, and
in the case of $N=4$ they obey the $SU(2)$ Clifford algebra.
These conditions put severe restrictions on the backgrounds that can
arise
as supersymmetric solutions.
For instance, in the absence of torsion, the metric should be Kahler
for $N=2$ and hyper--Kahler for $N=4$ \cite{ALFR}.

In order to proceed further we need to know the transformation
properties of
the complex structures under T--duality. If a complex structure
is independent of $X^0$, then under duality it transforms as
\cite{IKR,basfe}
\beq
 (\tilde F^\pm_I)_{0i}~ =~ \pm {(F^\pm_I)_{0i}\over G_{00}}~ ,~~~~~
 (\tilde F^\pm_I)_{ij}= (F^\pm_I)_{ij} ~ +~
{1\over G_{00}}
\left( (F^\pm_I)_{0i} Q^\pm_{j0} - (F^\pm_I)_{0j} Q^\pm_{i0}
\right)~ ,
\label{dualco}
\end{equation}
and hence defines a locally realized extended supersymmetry in the
dual model
as well.
If on the other hand it depends on $X^0$, then to obtain the right
transformation
one should replace $X^0$ by the non--local
expression (\ref{rednonl}) in the corresponding Kahler form;
in this case supersymmetry
will be realized non--locally in the dual picture \cite{basfe}.

As a consequence of having non-local realizations of supersymmetry,
many of the theorems established in the past
for 2-dim $\s$--models with $N=4$ extended supersymmetry are not
strictly valid, since they were implicilty relying on the assumption
that
the complex structures are local functions of the target space
variables. For instance, the transformed complex structures are not
covariantly constant when non--local realizations come into
play \cite{basfe,hassand}. Also,
the generators of the holonomy group will no longer commute with the
complex structures \cite{sferesto}, in contrast with the case of
locally realized $N=4$ (see, for instance, \cite{HSPB}).
In the absence of torsion, in particular,
non--local $N=4$ world--sheet supersymmetry does not imply that the
manifold is hyper--Kahler, as for local $N=4$
\cite{ALFR}. If any of these theorems is used as a
guiding principle,
some supersymmetries will appear to be lost in the effective field
theory
approach after duality, as the non--local ones cannot  be
distinguished anymore. Then, an apparent
paradox arises, as it was stated in \cite{bakasII}. If duality
provides an
equivalence between strings propagating in different spacetimes,
then how can it destroy other genuine symmetries such as
supersymmetry?
The resolution to this paradox, as we will see, is simply that
non--local world--sheet effects have to be taken into account
in order to obtain a consistent picture with supersymmetry
\cite{basfe}.

The criterion for having an $X^0$--dependent
complex structure is ultimately
connected with the way
it fits into representations of the isometry group, which in
our simple Abelian case is isomorphic to $U(1)$.
In $N=2$ extended world--sheet supersymmetry we may arrange for the
complex
structure to be a $U(1)$ singlet, and hence independent of $X^0$,
but for $N=4$ extended supersymmetry there are two distinct
possibilities: either all three complex structures are $U(1)$
singlets,
or one of them is a singlet and the other two form a doublet.
In the former case the original locally realized $N=4$
supersymmetry remains local,
and hence manifest after duality,
whereas in the latter it becomes non-manifest by breaking
to a local $N=2$ part, with the rest being realized non--locally.

A useful criterion
for distinguishing when Abelian T--duality preserves
manifest $N=4$ extended world--sheet
supersymmetry is \cite{sferesto}
\beq
\d_\mu Q^\mp_{0\nu} (F^\pm_I)^{\mu\nu} = 0~ ,~~~ I=1,2,3~ .
\label{critii}
\end{equation}
This condition is indeed satisfied by all three complex structures
if they are $U(1)$ singlets, whereas in the case of
a singlet plus a doublet it turns out that
the right hand side of (\ref{critii}), instead of vanishing, takes
the form
\cite{sferesto}
\beq
\d_\mu Q^\mp_{0\nu} (F^\pm_I)^{\mu\nu} =  {d\over 2}
\e_{I12} ~ .
\label{criev}
\end{equation}
Hence, the violation of (\ref{critii}) receives contribution
from the singlet complex structure, which here
and in the following will be labelled by $I=3$. Notice that
the right hand side of (\ref{criev}) is just
a constant, although not zero. Therefore, the validity of
(\ref{critii})
may be checked in practice by expanding locally the right hand side
around a reference target space point.
In particular, let us briefly examine the case of 4-dim
flat space with metric written in polar coordinates
\beq
ds^2 = d\rho^2 +  \rho^2 d\varphi^2 +   dx^2 + dy^2 ~.
\label{flat}
\end{equation}
This background has manifest $N=4$ world--sheet supersymmetry.
The complex structure that remains invariant
under $\varphi$--shifts by a constant is given by
\beq
F_3 ~= ~ \rho d\rho \wedge d\varphi ~ + ~ dx \wedge dy ~ .
\label{flcom}
\end{equation}
It is a matter of short computation to verify that (\ref{criev})
(for $I=3$ and $d=4$) is satisfied. Thus, T--duality with respect
to $\varphi$ will break manifest $N=4$ supersymmetry. The same is
true
for all spaces that can be locally approximated by (\ref{flat}), in
the vicinity of the fixed points of a Killing isometry.
For obvious reasons such Killing vectors fields, and the
corresponding string backgrounds, will be called rotational.

\subsection*{ Complex structures and parafermions }

Since the non--local world--sheet effects we have described
are characteristic of the string theoretical nature of
duality, we would like to see explicitly
how they become manifest in the conformal
field theory corresponding to a 2--dim $\s$--model.
For this reason, we choose to demonstrate the aspects of
our general framework in the special case of the 4--dim
semi--wormhole
solution, and its rotational dual background. An exact
description is available for this background
in terms of the $SU(2)\otimes U(1)$
WZW model and its cousins.

The semi--wormhole solution of 4--dim string theory provides
an exact conformal field theory background with $N=4$
world--sheet supersymmetry \cite{CHS,KAFK}. The $N=4$ superconformal
algebra can be locally realized in terms of four bosonic currents,
three
non--Abelian $SU(2)_{k}$ currents and one Abelian current with
background
charge $Q = \sqrt{2/(k+2)}$, so that the central charge is $\hat{c} =
4$. There are also four free--fermion superpartners, and the solution
is
described by the $SU(2)_{k} \otimes U(1)_{Q}$ supersymmetric WZW
model. It is convenient to parametrize the $SU(2)$ group element as
\beq
 g = e^{{i\over 2} (\tau - \psi) \s_3}
e^{i \varphi \s_2} e^{{i\over 2} (\tau + \psi) \s_3} ~ .
\label{su2}
\ee
Then the
background fields of this model are given by
\ba
 d s^2 & = & d\r^2
+ d\varphi^2 + \sin^2\varphi\ d\psi^2 + \cos^2\varphi\ d\tau^2 ~ ,
\nonumber  \\
 B_{\tau\psi} & = & \cos^2\varphi ~ , ~~~ \Phi = 2 \rho ~ ,
\label{abcworm}
\ea
where $\rho$ corresponds to the $U(1)$ factor.
For small $\varphi$ this background approaches
the flat space metric (\ref{flat}), with $\d/\d\psi$ playing the
role of the rotational Killing vector field.

The following analysis was essentially performed in \cite{basfe},
using a
different parametrization.
The three complex structures for the right sector are
\be
F^+_i = 2 d\rho \wedge \S_i -  \e_{ijk} \S_j \wedge \S_k ~ ,
\label{comri}
\ee
where the left invariant Maurer--Cartan forms of $SU(2)$,
$\S_i = -{i\over 2} Tr(g^{-1} d g \s_i)$, have been used:
\ba
\S_3 & = & \cos^2\varphi d\tau + \sin^2\varphi d\psi ~ ,
\nonumber \\
 \S_\pm & = &  \S_1 \pm i \S_2 =
e^{\pm i (\tau +\psi)} \left(\pm i d\varphi +
{1\over 2} \sin 2\varphi (d\tau- d\psi)\right) ~ .
\label{maur}
\ea
The complex structures for the left sector can be similarly
written down
\be
F^-_i = 2 d\rho \wedge \tilde \S_i -  \e_{ijk}
\tilde \S_j \wedge \tilde \S_k ~ ,
\label{comle}
\ee
where $\tilde \S_i = - {i\over 2} Tr(d g g^{-1} \s_i)$
are the right invariant Maurer--Cartan forms of $SU(2)$.
Their explicit expressions can be obtained from (\ref{maur})
letting $(\tau,\psi,\varphi)\to (-\tau,\psi,-\varphi)$,
up to an overall minus sign.
It is easy to see that under constant shifts
of $\psi$, $F^+_3$ is a singlet, whereas $F^+_1$ and $F^+_2$
form a doublet, and we may proceed similarly for the other chiral
sector.

T--duality with respect to the Killing vector
$\d/{\d \psi}$ gives the background \cite{RSS}
\ba
d\tilde s^2 & = & d\varphi^2 + \cot^2\varphi\ d\a^2
+ d\b^2 + d\r^2 ~ ,\nonumber \\
\tilde \Phi & =  & 2\r + \ln(\sin^2\varphi )~ ,
\label{dualworm}
\ea
with zero antisymmetric tensor,
which corresponds to the $(SU(2)_k/U(1)) \otimes U(1) \otimes U(1)_Q$
model;
to be precise we should make the redefinition of
variables $\a=\tilde \psi -{\tau\over 2}$ and
$\b=\tilde \psi +{\tau\over 2}$.
Note that although the torsion is zero, the Ricci tensor
is not zero due to the presence of a non--trivial dilaton. This means
that the
manifold is not hyper--Kahler. According to our previous discussion,
this would have been
a paradox of the effective theory if non--local world--sheet
effects were not properly taken into account in string theory.

The complex structure dual to $F_3^\pm$ is given by
\be
 \tilde F_3= d\rho \wedge d\b + \cot\varphi \ d\varphi \wedge d\a ~ ,
\label{dulftt}
\ee
and clearly defines local supersymmetries in both chiral sectors.
Since there is no torsion, there is
also no distinction between the + and the -- components.
However, the dual of the complex structures $F^\pm_{1,2}$ are
non--local
due to the explicit appearance of $\psi$ in (\ref{maur}), which after
duality is given by (cf. (\ref{rednonl}))
\be
\psi = \int (\cot^2\varphi \d_+ \a + \d_+ \b )d\s^+  -
(\cot^2\varphi \d_-\a + \d_- \b ) d\s^- ~ .
\label{duava}
\ee
We write the dual form of the complex structures as \cite{basfe}
\ba
\tilde F^+_1 & = & (d\r + i d\b)\wedge \Psi_+  +
(d\r - i d\b) \wedge \Psi_-  ~ , \nonumber \\
\tilde F^+_2 & = &  i (d\r + id\b) \wedge \Psi_+
 - i (d\r -i d\b) \wedge \Psi_- ~  ,
\label{comdu1}
\ea
for the right sector, and
\ba
\tilde F^-_1 & = & i (d\r -i d\b) \wedge \bar \Psi_+ -
i (d\r + i d\b)\wedge \bar\Psi_- ~ , \nonumber \\
\tilde F^-_2 & = & (d\r - id\b) \wedge  \bar\Psi_+
+ (d\r +i d\b)\wedge \bar\Psi_- ~ ,
\label{comdu2}
\ea
for the left sector. There appears to be a
distinction between the + and the -- components, although
the torsion vanishes, and this is a novel characteristic of the
non--local
realizations of supersymmetry \cite{sferesto}.

The parafermionic-type 1--forms are defined as
\ba
\Psi_\pm   & = & (d\varphi \pm i \cot\varphi \ d\a)
e^{\pm i(\b-\a + \psi)}~ ,
\nonumber \\
\bar\Psi_\pm & = & (d\varphi \mp i \cot\varphi \ d\a)
e^{\pm i(\a-\b + \psi)}~ ,
\label{paraff}
\ea
and they are non--local due to (\ref{duava}).
They have a natural decomposition in terms of $(1,0)$ and $(0,1)$
forms on the string world--sheet
\be
 \Psi_\pm =\Psi^{(1,0)}_\pm d\s^+ ~ + ~
\Psi^{(0,1)}_\pm d\s^- ~ ,~~~~
\bar\Psi_\pm = \bar\Psi^{(1,0)}_\pm d\s^+ ~ + ~
\bar\Psi^{(0,1)}_\pm d\s^- ~ .
\label{decof}
\ee
It can be easily verified using the classical equations of
motion for the model (\ref{dualworm}) that they satisfy the chiral
and
anti--chiral conservation laws
\be
\d_- \Psi^{(1,0)}_\pm=0~ ,~~~~~\d_+ \bar\Psi^{(0,1)}_\pm=0 ~.
\label{onss}
\ee
In this case, in fact, $\Psi^{(1,0)}_\pm$ and
$\bar \Psi^{(0,1)}_\pm$
are nothing but the classical parafermions \cite{BCR}
for the $SU(2)_k/U(1)$ coset, with
the field $\b$ providing the appropriate dressing to the full 4--dim
model
$(SU(2)_k/U(1)) \otimes U(1) \otimes U(1)_Q$. Thus, the original
local $N=4$
world--sheet supersymmetry breaks to a local part corresponding to
(\ref{dulftt}), and the rest is realized non--locally using the
non--local analogue of the complex structures (\ref{comdu1}),
(\ref{comdu2}).
At the (super)CFT level this manifests by
replacing the three non-Abelian $SU(2)_k$ currents with two
$SU(2)_k/U(1)$ parafermions
and one Abelian current in the realization of the $N=4$
superconformal
algebra \cite{KAFK}.

\subsection*{ Restoration of manifest supersymmetry }

Since Abelian T--duality acts as a $Z_2$ symmetry,
it is obvious that by dualizing the background (\ref{dualworm})
with respect to the
Killing vector field $\d/\d{\tilde \psi}$, where $\tilde \psi
={1\over 2} (\a +\b)$, we will recover back (\ref{abcworm}).
In the process, as it is expected, the ``dressed"
parafermions (\ref{paraff}) will become the usual
currents corresponding to the raising and lowering generators of
the $SU(2)$ algebra.
A suggestive way of thinking about it is that the duality
provides a mechanism for restoring manifest supersymmetry.
In the following we present a less trivial example based on the
results
reported in
\cite{sferesto}. The same example was first considered in \cite{AAB}
from a different, though equivalent point of view,
in connection with the restoration of manifest spacetime
supersymmetry.
The issues of spacetime supersymmetry will be discussed separately
in some detail in the next section.

Let us consider a 4--dim string background with
\ba
 ds^2 & = & d\varphi^2 + \cot^2{\varphi} dx^2 + d\r^2
+ R^2(\rho) dy^2~ ,
\nonumber \\
\Phi & = & \ln(\sin^2\varphi/R'(\rho))~ ,
\label{met}
\ea
and zero antisymmetric tensor. The function $R(\rho)$ is dynamical,
and it is
constrained by requiring 1--loop conformal invariance to
satisfy the differential equation
\be
R'=C_1 R^2 + C_2 ~ .
\label{condR}
\ee
The constants $C_1$, $C_2$ completely classify the
different solutions (see, for instance, \cite{KKdyn}).

It turns out that (\ref{met}) has $N=4$ extended supersymmetry
with the local $N=2$ part corresponding to the complex structure
\be
F_3= \cot \vphi\ d\vphi \wedge dx + R(\rho)\ d\rho \wedge dy~ ,
\label{colo}
\ee
whereas the rest is realized non--locally. In order to present the
analogue of the other two complex structures, it is convenient to
introduce parafermionic-type 1--forms similar to (\ref{paraff}).
They are given by
\ba
& \Psi^{(1)}_\pm & = (d\vphi \pm i \cot\vphi d x)~
e^{\pm i (-x + \th_1)}~ , ~~~~~ \bar \Psi^{(1)}_{\pm}  =
(d \vphi \mp  i \cot\vphi d x)~ e^{\pm i (x + \th_1)}~ ;
\nonumber \\
& \th_1 &\equiv \int \cot^2\vphi \d_+ x d\s^+
- \cot^2 \vphi \d_- x d\s^- ~ ,
\label{formse}
\ea
corresponding to the usual $SU(2)/U(1)$ classical parafermions,
and
\ba
& \Psi^{(2)}_\pm & = (d \rho \pm i R d y)~
e^{\pm i (c_2 y + \th_2)}~ ,~~~~ \bar \Psi^{(2)}_{\pm}  =
(d \rho \mp i R d y)~ e^{\pm i (-c_2 y + \th_2)} ~ ,
\nonumber \\
&\th_2& \equiv  \int (c_2- R') \d_+y  d\s^+
- (c_2-R')\d_-y d\s^-~ ,
\label{formee}
\ea
where $c_2$ is an arbitrary constant.
Then, the non--local complex structures assume the form
\cite{sferesto}
\be
 F_1^+ = \Psi^{(1)}_+ \wedge \Psi^{(2)}_+
+ \Psi^{(1)}_- \wedge \Psi^{(2)}_- ~  ,~~~~~
F_2^+ = i \Psi^{(1)}_+ \wedge \Psi^{(2)}_+
- i  \Psi^{(1)}_- \wedge \Psi^{(2)}_- ~ ,
\label{nonlocs}
\ee
and
\be
 F_1^- = \bar\Psi^{(1)}_+ \wedge \bar\Psi^{(2)}_+
+ \bar\Psi^{(1)}_- \wedge \bar\Psi^{(2)}_- ~ ,~~~~~
F_2^- = - i \bar\Psi^{(1)}_+ \wedge \bar\Psi^{(2)}_+  +
i \bar\Psi^{(1)}_- \wedge \bar\Psi^{(2)}_- ~ ,
\label{noloscc}
\ee
These define legitimate extended supersymmetries at the classical
level, and they satisfy the general equations for
having non--local complex structures \cite{sferesto}.

We now address the question: under what
circumstances is it possible, via a
duality transformation of (\ref{met}), to obtain a $\s$-model with
manifest
$N=4$ supersymmetry, so that all the complex structures are local
functions of the target space variables.
Although we may consider a general
$O(2,2)$ transformation on (\ref{met}), it turns out that only
for a particular choice of parameters we obtain the desired result,
which is described below.

Let us introduce the coordinate change
$ x= \psi -{\tau\over 2}$, $ y=\psi +{\tau\over 2 }$
and perform a T--duality transformation with respect to the symmetry
generated by the Killing vector $\d/{\d\psi}$.
The resulting background is
\ba
d\tilde s^2 & = & d\rho^2 + d\vphi^2 +{1\over 1 + R^2 \tan^2\vphi }
( \tan^2\vphi d\tilde\psi^2 + R^2  d\tau^2 )~ ,
\nonumber \\
\tilde B_{\tau\tilde\psi} & = & {1\over 1+ R^2 \tan^2\vphi}~ ,~~~
\tilde \Phi = \ln\left((\cos^2\vphi + R^2 \sin^2\vphi)/R' \right)~ .
\label{dubac}
\ea
For constant $R$ it can be interpreted as describing a continuous
line of
$J\bar J$-deformed $SU(2)\otimes U(1)_Q$ models \cite{defsu} with $R$
as the modulus. The value $R=1$ corresponds to the WZW point.
The model with $R$ as a function
of $\rho$ was considered in its Minkowski version
as a toy model for studying dynamical topology change in
string theory \cite{KKdyn}.
We have also denoted the dual variable of $\psi$ by $\tilde\psi$.
Their explicit relation is of course  non--local according to
(\ref{rednonl}). What is important here is the relation of the
corresponding world--sheet derivatives, which is found to be
\be
 \d_\pm \psi =  {1\over 1+R^2 \tan^2\vphi}
\left(\pm \tan^2\vphi \d_\pm  \tilde\psi
+{1\over 2} ((1-R^2 \tan^2\vphi) \d_\pm \tau)\right) ~ .
\label{reltt}
\ee
This will be used
in order to deduce the transformation of the functionals
$\th_1,\th_2$
defined in (\ref{formse}), (\ref{formee}), and then the
transformation of the
non--local complex structures (\ref{nonlocs}), (\ref{noloscc}).
The  phase factors
in $F^\pm_{1,2}$ that are responsible for their non--local nature are
\ba
 &&\th_1 + \th_2  \pm (c_2 y -x)  =
\pm (c_2 -1) \psi  \pm  \ha (c_2+1) \tau \nonumber \\
&& + \int \left( (c_2-R' + \cot^2\vphi)\d_+ \psi  +  \ha
(c_2-R' - \cot^2\vphi)\d_+\tau \right) d\s^+ - ( +\to -)~ .
\label{phas}
\ea
We find generically that the non--localities persist after
T--duality, except in a particular case where they
completely cancel out. This happens
if we choose $c_2=1$ and the function $R(\r)$ to satisfy
\be
R'=1-R^2 ~~~ \Rightarrow ~~~ R=1 ~~~  or~~~ \tanh\rho ~~~ or ~~~
 \coth \rho ~ .
\label{RRR}
\ee
Then, indeed, the phase factors (\ref{phas}) transform
just to $\tilde \psi \pm \tau$,
and the dual complex structures become local.

Among the three different solutions, the one with $R=1$ corresponds
to the WZW model for $SU(2)\otimes U(1)_Q$ given by (\ref{abcworm}).
Hence we see that a marginal deformation away
from the WZW point
($R=const. \neq 1$) leads to a loss of manifest $N=4$.
It can be shown \cite{sferesto} that this is a general statement
valid
for all WZW models based on quaternionic groups, with
$SU(2)\otimes U(1)$ being the most elementary example.

The solutions $R=\tanh \rho$ and $R=\coth\rho$ correspond to a new
4-dim background with manifest $N=4$ supersymmetry. In an
appropriate coordinate system it assumes the form \cite{sferesto}
\ba
&& ds^2 = e^{-\Phi}\ dx_i dx_i~ ,
{}~~~~  H_{ijk} = -  \e_{ijk}{}^l \d_l \Phi~ ,
\nonumber \\
&& \Phi =  \ha\ln\left((x_ix_i+1)^2 - 4  (x_3^2+x_4^2)\right)~ .
\label{wormd}
\ea
The metric is conformally flat with the conformal
factor satisfying the Laplace equation adapted to the flat space
metric,
i.e. $\d_i\d_i e^{-\Phi}=0$,
in agreement with a general theorem proved in \cite{CHS}.
The antisymmetric field strength solves the (anti)self--duality
conditions
of the dilaton--axion field and therefore the solution (\ref{wormd})
is an axionic--instanton.
The complex structures are
\ba
&& F^\pm_1 = e^{-\Phi}(- dx_1 \wedge dx_3 \pm dx_2\wedge dx_4)~ ,
\nonumber \\
&& F^\pm_2 = e^{-\Phi}(\pm dx_1 \wedge dx_4 + dx_2\wedge dx_3)~ ,
\nonumber \\
&& F^\pm_3 = e^{-\Phi}( dx_1 \wedge dx_2 \pm dx_3\wedge dx_4)~ .
\label{conew}
\ea
In fact these are the complex structures for all 4--dim axionic
instantons of the
form (\ref{wormd}) irrespectively of the particular dilaton field
$\Phi$
\cite{sferesto}, which is only needed for conformal invariance.
Geometrically the metric represents the throat of
a semi--wormhole. Notice that a
true semi--wormhole is obtained only by shifting $e^{-\Phi}$ by a
constant, since then asymptotically the space is Euclidean.
In our case this
corresponds to an $S$--duality transformation.
We also note that the metric has singularities not at a single point,
but in the ring $x_1=x_2=0$, $x_3^2+x_4^2=1$.
Therefore, the throat never becomes infinitely thin.
The background (\ref{wormd})
is a generalization of the $SU(2)\otimes U(1)$
semi--wormhole background \cite{CHS,KAFK}
to which our solution approaches for large values of the $x_i$'s.

It is important to emphasize that in trying to obtain a 2-dim
$\s$-model with
manifest $N=4$ supersymmetry via a duality transformation from
(\ref{met}),
at no point we required conformal invariance. The
entire treatment was completely classical and
the function $R(\rho)$ remained arbitrary.
Both (\ref{met}) and its dual (\ref{dubac}) have non--locally
realized $N=4$ supersymmetry at the classical level.
It turned out that the condition (\ref{RRR}) that led to manifest
$N=4$
supersymmetry for the dual model is also a particular case of
(\ref{condR}), with
$C_2=-C_1=1$, which guarantees 1--loop conformal invariance for both
models.
With these choices for $R(\r)$, the parafermionic
1--forms $\Psi^{(2)}_\pm$ and
$\bar \Psi^{(2)}_\pm$ correspond to the usual
classical non--compact parafermions of the $SL(2,R)/U(1)$ coset.
Then, the background (\ref{met}) corresponds to the direct product
$SU(2)/U(1)_k \otimes SL(2,R)_{-k-4}/U(1)$ and the
$N=4$ superconformal algebra is realized using as natural objects the
compact and non-compact parafermions \cite{KAFK}.
A similar CFT construction
for the new background (\ref{wormd}) remains an open problem.
Since $N=4$ is manifest at the classical level, it is expected
that the realization of the corresponding superconformal algebra
will be local, or
at least it will become local in the classical regime of large $k$.

We conclude this section with a few comments on the cases
where the isometry group of duality is non--Abelian
\cite{laos} (for earlier work on the subject see \cite{duearl}).
We have seen that assigning the complex structures to
representation
of the isometry group is a useful way to determine the fate of
supersymmetry under T--duality.
Let us consider the effect of
non--Abelian duality transformations on
$SO(3)$--invariant hyper--Kahler metrics.
In such cases the complex structures are either
$SO(3)$ singlets, thus remaining invariant under the non--Abelian
group action, or they form an $SO(3)$ triplet.
The Eguchi--Hanson metric corresponds
to the first case, while the Taub--NUT and the Atiyah--Hitchin
metrics to the second \cite{GIRU}.
It should be clear, then, that the dual version of the Eguchi--Hanson
instanton
with respect to $SO(3)$
will have an $N=4$ world--sheet supersymmetry locally realized.
On the other hand, applying non--Abelian $SO(3)$--duality
to the Taub--NUT and the Atiyah--Hitchin metrics will result in a
total loss
of all the locally realized extended world--sheet supersymmetries.
Instead, a non--local realization of supersymmetry will emerge in
such cases,
with three
non--local complex structures that satisfy the general conditions
of \cite{sferesto}. Also, in cases where non--Abelian duality is
performed on a WZW model, the non--local realizations can be
described in
terms of non--Abelian parafermions.
We hope to report some work along these lines elsewhere
\cite{workpro}.

\subsection*{ Duality and space--time supersymmetry}

The conventional definition of a string background with unbroken
target space
supersymmetry requires the existence of solutions of the Killing
spinor equations.
Then, the quantum field theory of the fluctuations around this vacuum
will be, at tree level, supersymmetric as well. The unbroken
supersymmetries are in one to one correspondence
with the independent solutions of these equations.

For concreteness consider $N=1$ supergravity in $d=10$
dimensions coupled to superYang--Mills, which is the
low energy approximation to heterotic string theory.
To simplify matters further, we will consider vacuum solutions with
the
gauge fields and their corresponding gluinos set equal to zero.
Then the Killing spinor equations are
\ba
&&\delta \Psi_\mu = \left(\d_\mu + {1\over 4} (\omega_\mu{}^{\a\b}
-\ha H_\mu{}^{\a\b} )  \gamma_{\a\b} \right) \xi =0~  ,
\nonumber \\
&&\delta \l = - \left(\gamma^\mu \d_\mu \Phi + {1\over 6}
H_{\mu\nu\l}
 \gamma^{\mu\nu\l} \right) \xi =0~ ,
\label{kilspi}
\ea
where $\Psi_\mu$ and $\l$ are the gravitino and
dilatino fields respectively.

As for the world--sheet supersymmetry,
the presence of rotational--type Killing vector fields
also results to a
breaking of manifest target space supersymmetry under Abelian
T--duality,
in the sense that
no Killing spinors exist in the dual background \cite{bakasII,BKO}.
As an example consider a 10-dim
background whose non-trivial part is given by the 4-dim
$SU(2) \otimes U(1)$ model (\ref{abcworm}). Then, the solution of
the Killing spinor equation is (\ref{kilspi})
\be
\left( \begin{array} {c}
\xi_+ \\ \xi_- \end{array} \right) = e^{-{i\over 2}\vphi \s_2}
e^{-{i\over 2}(\tau+\psi)\s_3}
\left( \begin{array} {c}
0 \\ \e_- \end{array} \right) ~ ,
\label{spisu2}
\ee
where $\e_-$ is the non--zero Weyl component of a constant
spinor. However, for the dual background (\ref{dualworm}) there are
no solutions of (\ref{kilspi}). This should be obvious from
(\ref{spisu2}),
as the Killing spinor has an explicit dependence on $\psi$ and
after duality it becomes non--local \cite{hassand}.
The crucial difference with the case of extended
world--sheet supersymmetry is that here the
lowest order effective field
theory is not enough at all to understand the fate of target space
supersymmetry under duality, since one has to generate the whole
supersymmetry algebra and not just its truncated part
corresponding to the Killing spinor equations.
Nevertheless, it is believed that T--duality does not destroy
target--space
supersymmetry in an appropriate string setting.
An approach to this problem has been made in
\cite{AAB}, using CFT concepts, and will will not be discussed
further.

We think that massive string modes play a crucial role in this
game,
as it is also apparent
by making contact with the work of Scherk and Schwarz \cite{JSJS}
on coordinate dependent compactifications. We propose a comparison
between these two problems. In Scherk--Schwarz,
dimensional reduction is performed not in the conventional
way, by assuming that all fields and transformation parameters are
independent of the Kaluza-Klein internal coordinates, but instead a
special factorized dependence on the compactified coordinates is
kept.
It turns out that some fields acquire mass in the
dimensionally reduced theory, although
all the fields are massless in the unreduced theory, leading to a
breaking of supersymmetry upon compactification. Recall that in our
case the Killing spinors also depend on the ``Kaluza--Klein''
coordinate $X^0$ when the isometries are of rotational type.
Since the Killing
spinors are the supersymmetry transformation parameters, we
expect by analogy with \cite{JSJS} that massive strings modes
should play a role not only in the dimensionally reduced theory,
but in the realization of the supersymmetry algebra
after duality as well,
which in turn renders the truncation to only the massless modes as
inconsistent. The question we are really raising here is how
to formulate correctly in a string framework the duality
transformations with respect to Scherk--Schwarz ``isometries",
and more generally with respect the any other conceivable
compactification of string theory. The results we have described
so far can be viewed as a preliminary exercise towards this
goal, which could bring many new ideas in the subject
with numerous physical applications that seemed paradoxical in
the effective field theory approach.

It is worth mentioning that the breaking of manifest target space
supersymmetry
occurs hand and hand with the breaking of local $N=4$ extended
world--sheet supersymmetry. However, although
in the latter case the $N=2$ part remains local,
manifest target space supersymmetry appears to be completely broken.
This can be attributed to the
relation between Killing spinors and
complex structures \cite{kilcom},
using $F_{\mu\nu} = \bar \xi \G_{\mu\nu} \xi $,
thus making possible to construct
local complex structures out of non--local Killing spinors.
Such an example is precisely the background (\ref{dualworm}), which
has manifest $N=2$ world--sheet supersymmetry, but
no manifest spacetime supersymmetry.

Finally, let us discuss briefly the
mechanism of restoring manifest supersymmetry from the
spacetime point of view. The starting point in \cite{AAB}
is also the background (\ref{dubac}), with a dynamical  modulus
$R(\rho)$.
Demanding that the dilatino equation in
(\ref{kilspi}) is satisfied leads to the first order equation
(\ref{RRR}). The gravitino equation is also satisfied, and the
explicit form
of the Killing spinor can be found \cite{sferesto}. In fact,
in the coordinate system (\ref{wormd}) the solution of
(\ref{kilspi}) is just the constant Weyl spinor.
Notice that contrary to the restoration of manifest world--sheet
supersymmetry, which required no quantum input at all (conformal
invariance was not even an issue there), restoring manifest target
space
supersymmetry requires the use of the dilaton
field $\Phi$, which is a 1--loop
quantum effect in the $\a'$--expansion.

The various issues we have discussed here on the relation between
duality
and supersymmetry may also be
relevant to string phenomenology in one way or another. If the
duality can break
or restore manifest supersymmetry, then this
phenomenon should certainly be taken into consideration in various
supersymmetry
breaking scenarios relying on the effective field theory approach.
``Apparently'' non--supersymmetric backgrounds, such as
(\ref{dualworm})
or (\ref{met}),
can qualify as vacuum solutions
to superstring theory when it is possible to restore manifest
supersymmetry
through non--local world--sheet effects (of the type we have
described)
at the string level.
In addition,
this raises the question whether various solutions of physical
interest in
black hole physics or cosmology could
have hidden supersymmetries in
a string
context. Since this possibility necessarily involves non--local
world--sheet effects,
it will be important to explore it further
in our effort to understand ways that string
theory
can resolve fundamental problems in physics, in particular the
quantum theory
of black holes.

\subsection*{ A speculation with black holes }

The various solutions of the lowest order effective 
theory provide 
only a semiclassical approximation to the exact 
conformal field 
theories that correspond to different string vacua. 
Although there 
exist many solution generating techniques to lowest order in 
${\alpha}^{\prime}$, our present technology with CFT is 
comparatively limited 
to only a few exact constructions. For example, the familiar 
Schwarzchild metric describing 4--dim black holes in general 
relativity is a solution of the $\beta$--function equations, 
and for sufficiently large black holes with 
$(Gm)^2 > {\alpha}^{\prime}$ the higher order corrections can 
be regarded as perturbation. An analysis of this problem was 
initially done in \cite{CaMyPe}. Finding the exact CFT of the
Schwarzchild black hole, however, remains an outstanding 
problem up to this date. The successful solution will certainly
help us to understand how string theory could cure some of 
the paradoxes associated with black hole evaporation. 
We would like to view these long standing problems as 
paradoxes of the low energy effective theory and there is 
hope to use supersymmetry for this purpose. 
Of course, we are thinking about non--local realizations 
of an underlying superconformal algebra, and we will present 
some speculations in that direction. 

It is true that ordinary 4--dim black holes have no 
manifest space--time supersymmetry, which is consistent 
with their property 
of having a non--zero temperature $T$ inversely 
proportional to their mass parameter $m$. Microscopic 
black holes are very hot, and it is for them that the 
${\alpha}^{\prime}$ expansion should not be trusted. 
Thus, stringy effects, if any, will manifest as 
paradoxes of the lowest order effective theory that has 
been used so far to describe black holes. 
Motivated from our
previous results, we may forget now T--duality, 
and twist things around asking the 
following question: is 
the unknown CFT of black holes superconformal, but with a 
non--local realization, in which case non--local world--sheet
effects might resolve the problems with their evaporation 
within string theory? The closest we can get to entertain 
this idea is by considering 
exact 4--dim CFT coset models built 
out of the 2--dim black hole coset $SL(2,R)/U(1)$. Our 
previous analysis 
explicitly demonstrates that in such toy models 
the lowest order geometry has no manifest space--time 
supersymmetry, but in the exact picture the parafermions 
provide the relevant realization of an 
underlying $N=4$ superconformal
algebra. For real black holes, we hope first to find a 
superconformal structure at the classical level, and when 
the exact CFT will be known to be able to promote it 
to an exact symmetry 
quantum mechanically, using non--local operators 
of the model analogous
to the parafermions.  

A possible way to proceed 
classically with the construction of non--local 
complex structures for the black hole geometry is motivated 
by the reinterpretation of T--duality as 
a non--local change of the target space variables. Recall at 
this point that for a large class of string backgrounds with 
Killing symmetries, T--duality is only one element of a 
bigger symmetry group of the $\beta$--function equations, 
also known as U--duality in its 
various discrete forms. A particularly interesting class of 
4--dim backgrounds is provided by geometries with two 
commuting Killing symmetries. These include flat space and 
the black hole geometry, because it is a static 
and axisymmetric 
configuration. An interesting aspect of this class of 
geometries is the presence of an infinite dimensional 
group that acts as a complete solution generating 
symmetry, connecting any other solution with two isometries
to the flat space geometry. This is the well known 
Geroch group in general relativity \cite{Geroch}, but its 
generalizations to string theory (including anti--symmetric 
tensor and dilaton fields) also exist \cite{mystuff}, 
unifying T with S--dualities as continuous groups of 
transformations. It is then natural to ask whether there is
a reinterpretation of this huge symmetry as non--local 
change of the target space variables, thus extending the
known result for T--duality. If this is not a formidable 
task, it will provide the right transformation rules of 
the complex structures of flat space. Since any other 
solution with two isometries can be generated from flat 
space in this fashion, non--local complex 
structures could be constructed according to the geometry. 

We do not expect, however, 
that the transformed complex structures of 
flat space will always satisfy the right integrability 
conditions for having an underlying superconformal 
algebra to each geometry. It will be very useful in 
this regard to find those elements of the Geroch group 
that can lead to the proper non--local complex structures 
for defining a superconformal theory at the classical 
level, though non--locally realized. The black hole 
background is very special in this line of investigation. 
It is a celebrated result of Belinski and Sakharov that 
black holes (including also an arbitrary 
NUT parameter that
determines their asymptotic behaviour) admit a solitonic
interpretation, namely they arise as a double--soliton 
solution from flat space, using the integrability of
the 2--dim reduced Ernst equation \cite{BeZa}. 
Thus, the Geroch group element that generates the black 
hole geometry from flat space is indeed special, and
hopefully good enough to produce the wanted 
superconformal structure.  

In conclusion, we have set up a framework, motivated from
ideas arising in the interplay between supersymmetry 
and duality, in order to explore the possibility of 
having an extended superconformal algebra for the 
black hole (and possibly many other backgrounds as well) 
in string theory. We hope to report some encouraging 
results in this direction in the future.  

\subsection*{ Acknowledgments }

We would like to thank the organizers for the warm
hospitality and the enjoyable atmosphere they have 
created during the workshop.


\end{document}